# Tuning the magnetic properties of Co nanoparticles by Pt capping

A. Ebbing,[1,a)] O. Hellwig,[2] L. Agudo,[3] G. Eggeler,[3] and O. Petracic[1,b)]

[1]Institute of Experimental Physics/Condensed Matter Physics, Ruhr-University Bochum,

44780 Bochum, Germany

[2]San Jose Research Center, Hitachi Global Storage Technologies, 3403 Yerba Buena Rd.,

San Jose, California 95135, USA

[3]Institute for Material Science, Ruhr-University Bochum, 44780 Bochum, Germany

We show that by capping Co nanoparticles with small amounts of Pt strong changes of the magnetic properties can be induced. The Co nanoparticles have a mean diameter of 2.7 nm. From magnetometry measurements we find that for zero and for small amounts of Pt (nominal thickness $t_{Pt} < 0.7$ nm) the nanoparticles behave superparamagnetic like. With increasing $t_{Pt}$ the blocking temperature is enhanced from 16 up to 108 K. Capping with Pd yields comparable results. However, for values $t_{Pt} > 1$ nm a strongly coupled state is encountered resembling a ferromagnet with a $T_c \sim 400$ K.

Presently many efforts are undertaken to enhance the thermal stability of magnetic nanoparticles (NPs) for e.g. magnetic data storage media.[1-3] Various strategies have been proposed of how to achieve very high anisotropies and hence to 'beat the superparamagnetic limit'.[1,4] One route is to use Co/Pt or Fe/Pt multilayers or FePt and CoPt $L1_0$ phases.[5,6] Furthermore such NPs are expected to show also a perpendicular magnetic anisotropy (PMA) with respect to the plane of the recording medium to enable perpendicular recording.[7] From thin film studies it is known that Co/Pt multilayers show PMA with relatively large anisotropy values due to interfacial hybridization via the orbital moment of the Co surface

---
a) Electronic mail: Astrid.Ebbing@ruhr-uni-bochum.de
b) Electronic mail: Oleg.Petracic@ruhr-uni-bochum.de

atoms.[8,9] Therefore, we have investigated the magnetic properties of Co NPs, which have been capped with Pt in order to systematically influence the anisotropy of the NPs.

Recently, Bartolomé et. al. have shown[10] that capping Co NPs with a Pt layer of a nominal thickness of $t_{Pt} > 1.5$ nm leads to an increase of the superparamagnetic (SPM) blocking temperature. Furthermore, increased capping leads to a coupled state of the isolated NPs termed 'correlated superspin glass system'. However, in this study the NPs suffered from alloying during preparation due to Co/Pt interdiffusion.

Here we discuss the effect of capping the Co NPs with various and much smaller amounts of Pt. The samples were prepared at room temperature by ion beam sputtering at base pressures better than $5 \cdot 10^{-9}$ mbar using highly purified Ar gas. After sputtering the amorphous $Al_2O_3$ buffer layer of 3.4 nm thickness from an $Al_2O_3$ target onto Si substrates a cobalt layer of *nominal* thickness $t_{Co} = 0.66$ nm was sputtered from a cobalt target under a constant oblique deposition angle of 30° with respect to the surface normal. Due to extreme Volmer-Weber growth the Co forms isolated and nearly spherical particles.[11-14] These particles were then capped by sputtering a Pt layer with various *nominal* thicknesses $0 \leq t_{Pt} \leq 1.58$ nm again under a constant deposition angle of 30°. Finally, an alumina layer with a thickness of 3.4 nm was sputtered under constant rotation of the substrate to embed and to protect the NPs from oxidation. In addition, reference samples were prepared by magnetron sputtering with $t_{Co} = 0.69$ nm using Pd instead of Pt with $t_{Pd} = 0$, 0.24 and 0.70 nm capped with 4 nm Ta instead of alumina.

To study the magnetic properties of the system, we performed magnetometric measurements using a superconducting quantum interference device (SQUID) magnetometer (MPMS, Quantum Design). The magnetization data is hereby normalized to the deposited volume of cobalt. To study the morphology, transmission electron microscopy (TEM) and scanning TEM (STEM) images were taken using an Analytical FEG-TEM TECNAI F20 S-Twin instrument, working at 200kV. The TEM samples were prepared using KBr crystals as substrates instead of Si. The different substrate is not expected to alter the sample morphology, because of the amorphous alumina as buffer layer. After the deposition process, the crystals were dissolved in water and the film fragments were placed on Cu TEM grids.

Fig. 1 shows STEM images of samples with 0, 0.53 and 1.40 nm Pt. Without Pt capping (Fig. 1A), the Co particles are isolated and have an average diameter of 2.7 nm at average distances of 4.2 nm. Fig. 1B shows particles capped with 0.53 nm Pt. Here, the NPs are randomly connected via narrow bridges of Pt. Capping with 1.40 nm Pt (Fig. 1C) leads to complete percolation between the particles.

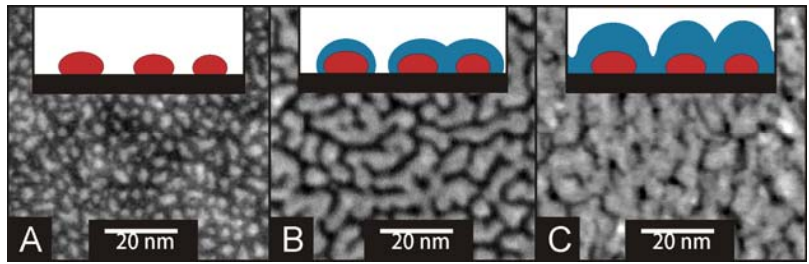

Fig. 1. STEM images on systems with constant $t_{Co}$ = 0.66 nm yielding Co NPs with an average diameter of 2.7 nm. The Pt coverage is varied, i.e. $t_{Pt}$ = 0 (A), 0.53 (B) and 1.40 nm (C). Hereby, dark regions correspond to the alumina background and bright regions correspond to Co or Co/Pt, respectively. The insets show schematic crossections.

Fig. 2 shows the magnetization $M$ vs. $T$ of the samples for constant $t_{Co}$ = 0.66 nm and various $t_{Pt}$ = 0, ..., 0.53 nm. The applied field and measurement axis were in-plane. The samples were first cooled down to 5 K in zero field and then a field of 20 Oe was applied. Then, the 'zero field cooled' (ZFC) magnetization curve was measured upon warming. Subsequently, in the same applied field, the 'field cooled' (FC) curve was recorded during cooling. The samples show superparamagnetic (SPM) type behavior as found from the typical shape of the ZFC curve with a peak at the blocking temperature, $T_B$, and a splitting of the ZFC-FC curves near $T_B$.[15-17] The inter-particle magnetostatic interactions are not negligible as evidenced from measurements of the so-called memory effect (data not shown). This is considered to be a fingerprint of collective superspin glass (SSG) behavior.[16,17]

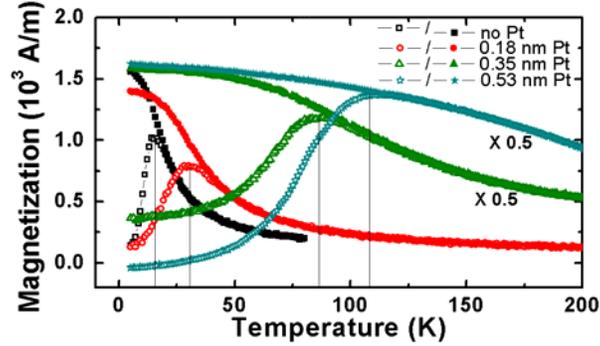

Fig. 2. ZFC (open symbols) and FC (filled symbols) in-plane measurements of $M$ vs. $T$ at the same field of $H = 20$ Oe for samples with constant $t_{Co} = 0.66$ nm and various $t_{Pt} = 0, ..., 0.53$ nm as indicated in the legend. The curves for 0.35 and 0.53 nm Pt were scaled down by a factor of 0.5 for better clarity.

Comparing the ZFC curves for various Pt thicknesses we observe an enhancement of the blocking temperature from 16 K (uncapped) to 108 K (0.53 nm Pt). Since the phenomenon of SPM blocking is related to magnetic anisotropy,[17-19] this enhancement can be attributed to an increased particle anisotropy. It was demonstrated[20] that for thin Co/Pt layers the 5d orbitals of the Pt are magnetically polarized. In the case of NPs this polarization may enhance the effective anisotropy of the single particles. In addition one can assume that random Pt bridges between NPs couple them to clusters of NPs (see Fig. 1B), hence increasing the effective magnetic volume and thus increasing the blocking temperature.

The effect of Pd capping yields analoguous results (data not shown). Here also an increase of the blocking temperature occurs, which can be attributed to the polarized 4d orbitals of Pd. The inset of Fig. 4 shows the dependence of $T_B$ on $t_{Pt/Pd}$. One clearly observes a monotonic increase of $T_B$ with increasing nominal thickness for $t_{Pt/Pd} < 0.7$ nm.

One should note that the amplitude of the magnetization signal (i.e. after normalization to the Co material volume) does *not* decrease with increasing Pt coverage as was reported in Ref. 10. There, the decrease was interpreted in terms of Co-Pt interdiffusion. In our case the amplitude does not seem to have a systematic dependence upon $t_{Pt}$. We assume that negligible interdiffusion occurs in our case. This is likely to be due to the nature of the ion beam sputtering process, i.e. relatively cold atoms impacting onto the substrate and a relatively large distance between target and substrate ($\approx$ 30 cm).

Magnetization hysteresis loops in the in-plane geometry show regular S-shaped open loops for $T < T_B$ and closed loops for $T > T_B$. As an example, $M(H)$ curves are shown for the case of $t_{Pt} = 0.35$ nm in Fig. 3 for $T = 5$ and 300 K. In contrast, corresponding out-of-plane curves show a shallower S-shape (data not shown). This is the case for all investigated samples indicating that the easy axis lies in-plane. Obviously the Pt capping has not the effect of establishing PMA. This might be due to the asymmetry of the sandwich structure $Al_2O_3$/Co/Pt, which has been reported to show PMA in thin film multilayers only after certain annealing conditions.[21]

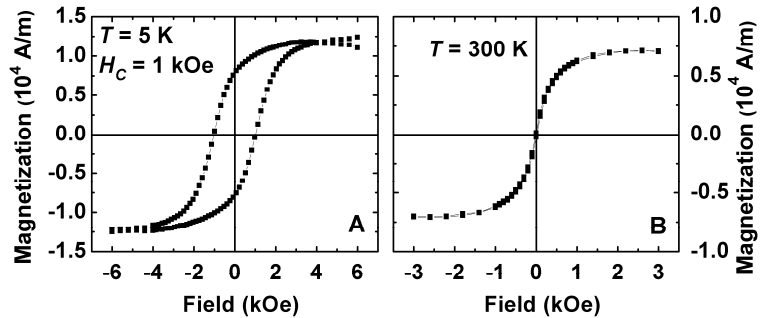

Fig. 3. $M$ vs. $H$ hysteresis loops for the sample with $t_{Pt} = 0.35$ nm measured in-plane at 5 K (A) and 300 K (B), respectively.

A further increase of the amount of Pt, i.e. 1.05 nm $\leq t_{Pt} \leq$ 1.58 nm, leads to a significant change of the magnetic behavior. Fig. 4 shows, as an example, the ZFC and FC magnetization of the sample capped with 1.40 nm Pt. Initially a field of -1 kOe was applied at 350 K. To allow magnetic relaxation, the sample was subsequently kept at 350 K for 30 min at zero field and then cooled to 5 K in zero field.

The ZFC/FC curves are typical of a ferromagnet (FM). I.e. the ZFC curve shows a sharp switching at 210 K, which corresponds to the temperature, where the coercivity of the hysteresis loop matches the applied field of 20 Oe. Furthermore, the FC curve shows the typical order parameter behavior of a FM. Measurements of other samples with $t_{Pt} = 1.05$, 1.23 and 1.58 nm show similar results (data not shown). Since the $M(T)$ measured at relatively low fields corresponds approximately to the FM order parameter, $M_s$, one can obtain the expected Curie temperature of this system using the semi-empirical formula given in Ref. 22, i.e. $M_s = M_s(0)\left[1 - s(T/T_c)^{3/2} - (1-s)(T/T_c)^p\right]^\beta$, where $0 < s < 5/2$ and $p > 3/2$ are semi-empirical fit parameters and $\beta$ the critical exponent of the order parameter.

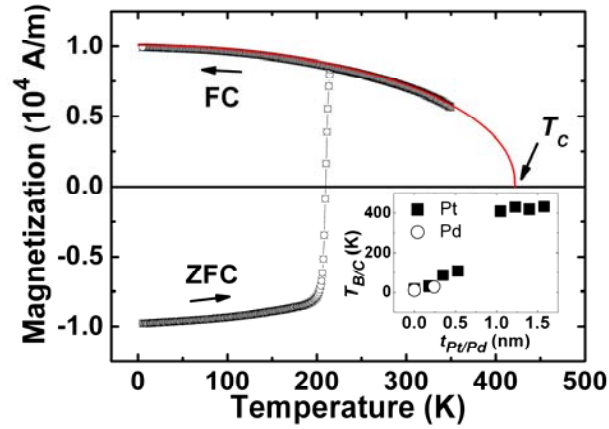

Fig. 4. ZFC and FC in-plane measurement of $M$ vs. $T$ at $H = 20$ Oe for the sample with $t_{Pt} = 1.40$ nm. The line extending beyond the data points is the fit to the FC data as described in the text with the resulting Curie temperature indicated by an arrow. The inset shows the dependence of $T_B$ or $T_c$, respectively, vs. the nominal thickness of Pt/Pd.

Fits to the data for samples with $t_{Pt} = 1.05$, 1.23, 1.40 and 1.58 nm yield Curie temperatures $T_c = 409 \pm 5$, $429 \pm 17$, $418 \pm 1$ and $431 \pm 2$ K, respectively, $s = 0.10 \pm 0.1$, $0.07 \pm 0.05$, 0 and $0.028 \pm 0.004$, respectively, $p = 1.91 \pm 0.11$, $1.80 \pm 0.10$, $1.88 \pm 0.01$ and $1.64 \pm 0.01$,

respectively and $\beta$ = 0.37 ± 0.01, 0.43 ± 0.06, 0.44 ± 0.01 and 0.44 ± 0.01, respectively. Hereby the errors were determined from fits to multiple different measurement curves recorded under same conditions on each sample.

The inset of Fig. 4 shows the dependence of $T_c$ on $t_{Pt}$. The $T_c$ values are similar being ~400 K and significantly higher than the blocking temperature as observed in the case for less Pt. The critical exponent for $t_{Pt}$ = 1.05 nm, i.e. $\beta$ = 0.37 corresponds approximately to the theoretical 3d Heisenberg value of 0.38,[23] whereas for larger $t_{Pt}$ it approaches the theoretical mean field value of 1/2. This can be explained by a long-range nature of the inter-particle interactions via the Pt bridges. This is consistent with the relatively small value for the shape-parameter $s$ ~ 0.1. A small value indicates long-range ferromagnetic exchange interactions.[24]

The observed FM like behavior is either due to single NPs, which behave as stable FM nanomagnets or due to an inter-particle coupled state. This can be distinguished from magnetization hysteresis loops. Fig. 5 shows $M(H)$ loops of the sample with 1.40 nm Pt at 5 K and 300 K, respectively. One clearly finds for the in-plane case FM-like sharp switching for both temperatures. Comparison with the much shallower out-of-plane loops (data not shown) indicates that the easy axis lies in-plane as in the cases discussed above.

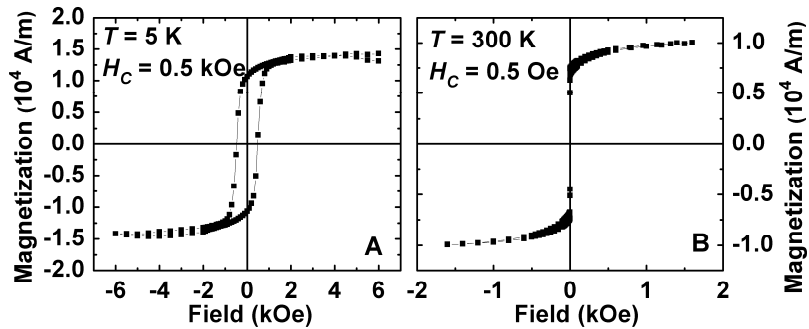

Fig. 5. $M$ vs. $H$ hysteresis loops for the sample with $t_{Pt}$ = 1.40 nm measured in-plane at 5 K (A) and 300 K (B), respectively.

Moreover, the sharp switching together with the relatively large squareness of the loops suggests that the FM like state is not due to single FM NP behavior. This is because an ensemble of FM NPs would yield a rounded loop after averaging over a random distribution of anisotropy axes. The sharp switching rather suggests reversal by domain wall motion in a collective state of many coupled NPs. The coupling must be due to the Pt bridges between Co NPs. We assume that the FM coupling is a consequence of the polarization of the Pt. It is less likely due to a RKKY interaction mediated by the Pt, since the average clearance between NPs is relatively large, ≈1.5 nm.

Measurements on samples with 1.05, 1.23 and 1.58 nm Pt yield completely similar results (data not shown). The remanent magnetization remains approximately equal, whereas the coercivity decreases for increasing Pt amount. This is likely to be due to less pinning because of increased inter-particle coupling. In order to distinguish this system from a regular exchange coupled FM and from a purely dipolarly coupled 'superferromagnet',[14,17] we term this state 'correlated granular ferromagnet' (CFM). The results from the Pt series are summarized in the inset of Fig. 4. The intermediate region, $0.7$ nm $< t_{Pt/Pd} < 1$ nm, is a transition region yielding a complex behavior eluding presently any explanation and will be reported in a forthcoming work.

In conclusion, we prepared Co NPs with a mean diameter of 2.7 nm by ion beam sputtering of cobalt onto an alumina buffer layer. When capping these NPs by a relatively small amount of Pt or Pd, i.e. $0 < t_{Pt/Pd} < 0.7$ nm nominal thickness, one obtains SPM-type (i.e. SSG) behavior. An increase in $t_{Pt/Pd}$ yields a monotonic increase in the blocking temperature. TEM studies reveal isolated particles, which are increasingly connected by Pt/Pd bridges. For large Pt nominal thicknesses, i.e. $1.05$ nm $< t_{Pt} < 1.58$ nm, we encounter a 'correlated granular ferromagnet' (CFM), where FM type of coupling between Co NPs exists. The coupling is likely to be mediated by a percolated network of Pt bridges.

We thank Hartmut Zabel for valuable discussions and support. Moreover, financial support by the Materials Research Department of the Ruhr-University Bochum is acknowledged.

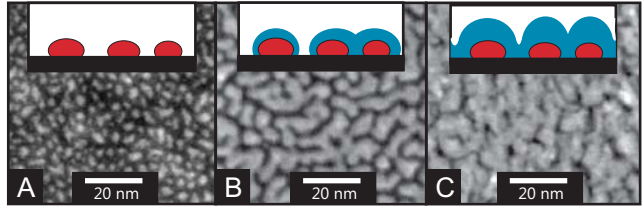

Figure 1

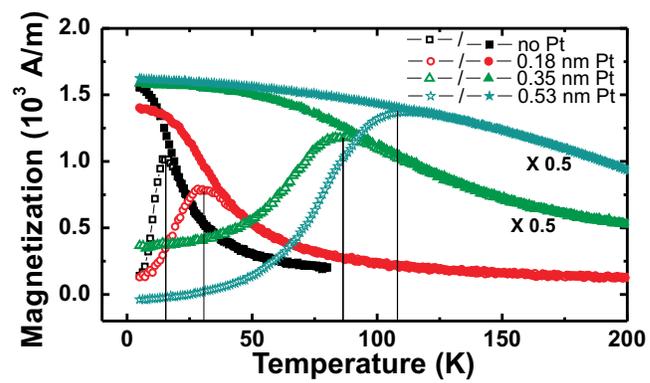

Figure 2

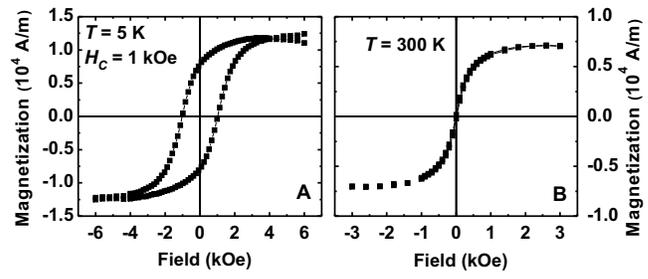

Figure 3

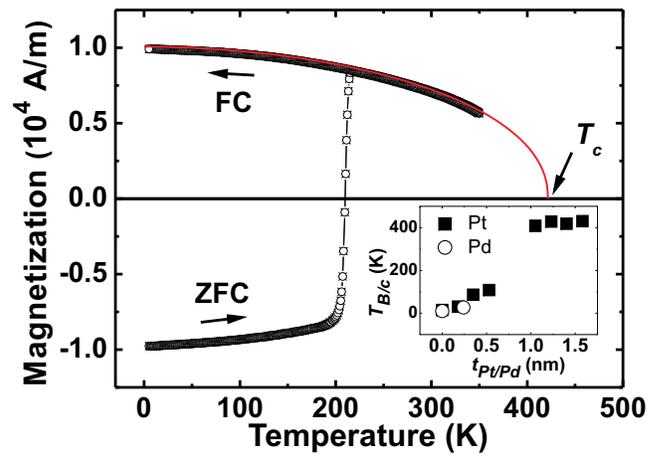

Figure 4

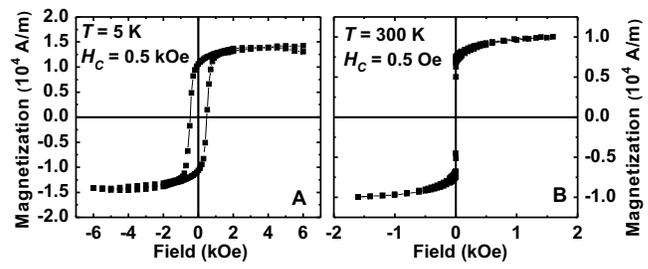

Figure 5